\documentstyle[floats,aps]{revtex} 

\input epsf         
\epsfverbosetrue    
\def\postscript#1{\begin{center}\leavevmode
\hbox{\epsfxsize=0.95\columnwidth\epsfbox{#1}}\end{center}}

\begin{document}

\twocolumn[\hsize\textwidth\columnwidth\hsize\csname@twocolumnfalse%
\endcsname

\draft

\title{Effect of anisotropy on
universal transport in unconventional superconductors}

\author{W. C. Wu}
\address{Department of Physics, National Taiwan Normal University,
Taipei 11718, Taiwan}

\author{D. Branch and J. P. Carbotte}
\address{Department of Physics and Astronomy, McMaster University\\
Hamilton, Ontario, Canada L8S 4M1}
 
\date{\today}
\maketitle

\begin{abstract}
We investigate the universal electronic transport
for a mixed $d_{x^2-y^2}$+$s$-wave superconductor in the presence of
an anisotropic elliptical Fermi surface.
Similar to the universal low-temperature transport predicted
in a $d_{x^2-y^2}$-wave superconductor with a circular Fermi surface,
anisotropic universal features are found in the
low-temperature microwave conductivity, and thermal conductivity 
in the anisotropic system.  The effects of anisotropy on the 
penetration depth, impurity induced $T_c$ suppression, and the
zero-frequency density of states are also considered.
While a small amount of anisotropy
can lead to a strong suppression of the effective scattering rate 
and hence the density of states at zero frequency, experimental data 
suggests that large effects are 
restored by a negative $s$-component gap admixture.
\end{abstract}

\vskip 0.2 true in
]

\narrowtext

\section{Introduction}
\label{sec:introduction}

It is well known that in high-temperature superconductors, owing to
their layered nature and their reduced dimensionality, there exits significant
anisotropy between various properties parallel and perpendicular to 
the copper oxide layers \cite{CG94}. 
In some materials such as YBCO, anisotropy is also 
seen between properties along the the $a$ and $b$ axes. 
These in-plane anisotropies are due to the presence of a CuO
chain layer in addition to the CuO$_2$ plane layers. Thus the overall
band structure parallel to the layer is orthorhombic instead of tetragonal. 
On the other hand, 
the symmetry of the superconducting gap remains controversial in 
high-$T_c$ compounds. It is generally believed to be $d_{x^2-y^2}$-wave,
while many suggest that there is an additional $s$-wave component
\cite{Sun,Kouznetsov,Kleiner,Aubin}.

For a superconductor with order parameter with nodes
on the Fermi surface, it is known that impurity scattering can
lead to a finite density of quasiparticle states at zero energy
(gapless excitations). As a consequence, 
while the effective impurity scattering rate $\gamma$ is quite different
for different impurity concentrations and different scattering limits, 
the in-plane microwave conductivity saturates at low temperatures 
and is independent of $\gamma$ (or the impurity concentration).
This universal transport phenomenon was first predicted by Lee \cite{Lee93}
for a $d_{x^2-y^2}$-wave superconductor.
Later Graf {\em et al.} \cite{GYSR96} found a similar
universality in the thermal conductivity which is
related to the microwave conductivity via the Wiedemann-Franz law.
Recent thermal conductivity experiments done by Taillefer {\em et al.}
\cite{TLGBA97} have confirmed these universal features.
More recently, Wu and Carbotte \cite{Wu98-1} demonstrated how one can 
also study these universal features by doing channel-dependent Raman 
scattering experiments.
It was found, in a $d_{x^2-y^2}$-wave superconductor,
that the low-temperature slope of
the Raman intensity at zero frequency are universal in the
$A_{1g}$ and $B_{2g}$ channels, but strongly dependent
on the scattering rate ($\sim \gamma^2$) in the $B_{1g}$ channel \cite{Wu98-1}.

The saturation of the low-temperature
transport in an unconventional superconductor is
a result of a cancellation between the value of the impurity-induced
density of states at zero energy and the
quasiparticle relaxation lifetime. It thus arises only if
there exists nodes in the order parameter on the Fermi surface.
In this paper, we consider the orthorhombic nature of the 
crystal structure in YBCO and 
explore whether an anisotropic universality exists in such a system.
For simplicity, we study a system with
an elliptical Fermi surface (assuming that the effective masses are different
along $x$ and $y$ directions). We have simultaneously considered
a $d_{x^2-y^2}$-wave gap with an additional $s$-wave component.
We investigate various quantities, including the in-plane low-temperature 
microwave conductivity, thermal conductivity, and the
London penetration depth. 
It will be shown that a small amount of anisotropy (the combined
contribution from energy band anisotropy and gap symmetry anisotropy)
can lead to a drastic
change in the effective scattering rate in the superconducting state.
As a consequence, the system exhibits
an anisotropic universal feature in various transport quantities.
More interestingly, experimental data \cite{Basov,Taillefer}
seems suggesting that the effect of the large energy band  
anisotropy is largely compensated for by a {\em negative} $s$-wave
component distortion to the $d_{x^2-y^2}$-wave gap. 

The content of this paper is as follows. 
In Sec.~\ref{sec:formalism}, we establish the Kubo
formalism needed for subsequent calculations.
In Sec.~\ref{sec:model}, we present our starting anisotropic model for a
$d_{x^2-y^2}$+$s$-wave superconductor with an orthorhombic elliptical
Fermi surface. In Sec.~\ref{sec:transport}, the anisotropic universal
microwave conductivity, anisotropic magnetic penetration depth,
and anisotropic universal thermal conductivity are presented.  
In Sec.~\ref{sec:self-energy}, we discuss how the effective 
scattering rate and hence the density of states at zero frequency
are modified in such an anisotropic system.
The effect of anisotropy on the 
impurity-induced $T_c$ suppression is also considered.
A short conclusion is given in Sec.~\ref{sec:conclusions}.

\section{Basic Formalism}
\label{sec:formalism}

We derive first the formalism
for the optical conductivity in details. The microwave conductivity and
the London penetration follow and are 
given by the real and imaginary parts
of the optical conductivity in the limit of zero frequency. 
Later we will briefly derive
the formalism for the thermal conductivity which
shares a great similarity with that of the optical conductivity.

The optical conductivity is given by \cite{Nam}

\begin{eqnarray}
&&\sigma_{\mu \nu}(\Omega)=\nonumber\\
&&{i\over \Omega}\left[
K_{\mu \nu}({\bf q}\rightarrow 0,i\nu_n\rightarrow
\Omega+i\delta)-K^{(n)}_{\mu \nu}({\bf q}\rightarrow 0,0)\right],
\label{eq:j.response}
\end{eqnarray}
where $K_{\mu \nu}$ is the paramagnetic
Kubo function given by

\begin{eqnarray}
&&K_{\mu \nu}({\bf q},i\nu_n)= {e^2 T\over 2} \sum_{{\bf k},\omega_{n}} 
\label{eq:K.def} \\
&&{\rm Tr} \left[\hat{\gamma}_\mu({\bf k+{q\over 2}})
\hat{G}({\bf k},i\omega_{n}+i\nu_n))\hat{\gamma}_\nu({\bf k+{q\over 2}})
\hat{G}({\bf k+q},i\omega_{n})\right],\nonumber
\end{eqnarray}
with Tr denoting a trace.  In Eq.~(\ref{eq:j.response}),
the second term $K^{(n)}_{\mu \nu}$ 
is calculated in the normal state and
with a minus sign corresponds to a diamagnetic response. 
The current vertex in Eq.~(\ref{eq:K.def}) is given by

\begin{eqnarray}
\hat{\gamma}_\mu({\bf k})={1\over \hbar}{\partial\xi_{\bf k}\over
\partial k_\mu} \hat{\tau}_0, 
\label{eq:j.vertex}
\end{eqnarray}
with $\hat{\tau}_i$'s the Pauli matrices and
$\xi_{\bf k}$ the electronic dispersion relation of
the superconducting layer.
In Eq.~(\ref{eq:K.def}), we have ignored the contribution
to the vertex corrections
due to the impurity potentials and superconducting
two-particle pairing interactions.
For isotropic impurity scattering,
it is sufficient to use the bubble diagram at small ${\bf q}$,
while the inclusion of the pairing interaction vertex correction is
negligible at the low frequencies of interest.
However, the effect of impurities is fully included in the
single-particle matrix Green's function $\hat{G}$ in Eq.~(\ref{eq:K.def}).

In terms of the particle-hole space, the single-particle 
matrix Green's function is given by

\begin{eqnarray}
\hat{G}({\bf k},i\omega_n)= {i\tilde{\omega}_n\hat{\tau}_0+
\tilde{\xi}_{\bf k}\hat{\tau}_3+\tilde{\Delta}_{\bf k}\hat{\tau}_1
\over (i\tilde{\omega}_n)^2-
\tilde{\xi}^2_{\bf k}-\tilde{\Delta}^2_{\bf k}}
\label{eq:G}
\end{eqnarray}
where $\tilde{\omega}_n$, $\tilde{\xi}_{\bf k}$, and $\tilde{\Delta}_{\bf k}$
are the impurity-renormalized Matsubara frequencies,
electron energy spectrum,
and gap. $\hat{G}$ is related to the noninteracting
Green's function $\hat{G}_0^{-1}({\bf k},i\omega_n)= i{\omega}_n\hat{\tau}_0-
{\xi}_{\bf k}\hat{\tau}_3-{\Delta}_{\bf k}\hat{\tau}_1$
via the Dyson's equation 

\begin{eqnarray}
\hat{G}^{-1}({\bf k},i\omega_n)
=\hat{G}_0^{-1}({\bf k},i\omega_n)-\hat{\Sigma}({\bf k},i\omega_n).
\label{eq:}
\end{eqnarray}
We shall solve the self-energy
$\hat{\Sigma}$ due to the impurity scattering.
By expanding $\displaystyle \hat{\Sigma}({\bf k},i\omega_n)\equiv \sum_{\alpha}
\Sigma_\alpha ({\bf k},i\omega_n) \hat{\tau}_\alpha$ ($\alpha=0,1,3$),
one finds $i\tilde{\omega}_n=i\omega_n-\Sigma_0$,
$\tilde{\xi}_{\bf k}={\xi}_{\bf k}+\Sigma_3$,
and  $\tilde{\Delta}_{\bf k}=\Delta_{\bf k}+\Sigma_1$.
Employing the usual $T$-matrix approximation, the self-energy is then
given by $\hat{\Sigma}({\bf k},i\omega_n)=n_i
\hat{T}({\bf k,k},i\omega_n)$, where
$n_i$ is the impurity density and
 
\begin{eqnarray}
&&\hat{T}({\bf k,k^\prime},i\omega_n)=
v_i({\bf k,k^\prime})\hat{\tau}_3\nonumber\\
&&+\sum_{\bf k^{\prime\prime}}
v_i({\bf k,k^{\prime\prime}})\hat{\tau}_3\hat{G}({\bf k^{\prime\prime}},
i\omega_n)
\hat{T}({\bf k^{\prime\prime},k^\prime},i\omega_n).
\label{eq:t.matrix}
\end{eqnarray}
Here $v_i({\bf k,k^\prime})\equiv
\langle{\bf k^\prime}|v_i|{\bf k}\rangle$ is the impurity potential.
If we consider only isotropic impurity scattering
[$v_i({\bf k,k^\prime})= v_i$], the $T$-matrix in (\ref{eq:t.matrix}) is
left only with frequency dependence and can be solved to get

\begin{eqnarray}
\hat{T}(i\omega_n)= \left[1-v_i\hat{\tau}_3
\hat{\overline{G}}(i\omega_n)\right]^{-1} v_i\hat{\tau}_3
\label{eq:T.simplified}
\end{eqnarray}
with the integrated Green's function $\hat{\overline{G}}(i\omega_n)\equiv
\sum_{\bf k}\hat{G}({\bf k}, i\omega_n)$.
One can expand $\displaystyle
\hat{\overline{G}}(i\omega_n)=\sum_{\alpha} G_\alpha(i\omega_n)
\hat{\tau}_\alpha$ ($\alpha=0,1,3$) with
$G_\alpha(i\omega_n)\equiv 1/2 \sum_{\bf k}{\rm Tr}[\hat{\tau}_\alpha
\hat{G}({\bf k},i\omega_n)]$.
For a superconductor with particle-hole symmetry,
$G_3(i\omega_n)=0$ and one obtains

\begin{eqnarray}
\Sigma_0(i\omega_n)&=&{n_i G_0(i\omega_n)\over c^2- G_0^2(i\omega_n)
 + G_1^2(i\omega_n)} \nonumber\\
\Sigma_1(i\omega_n)&=&{-n_i G_1(i\omega_n)\over c^2- G_0^2(i\omega_n)
 + G_1^2(i\omega_n)}\nonumber\\
\Sigma_3(i\omega_n)&=&{c n_i \over c^2- G_0^2(i\omega_n)  + G_1^2(i\omega_n)},
\label{eq:Sigmas.1}
\end{eqnarray}
where $c^2\equiv 1/v_i^2$. 
The effect of $\Sigma_3$ is usually absorbed
into the chemical potential and consequently
$\tilde{\xi}_{\bf k}\equiv \xi_{\bf k}$ or $\Sigma_3\equiv 0$.
It is noted that for a system with a cylindrical Fermi surface and
a pure $d$-wave gap,
$G_1=0$ and hence $\Sigma_1=0$ which gives no
renormalization effect on the gap ($\tilde{\Delta}_{\bf k}=\Delta_{\bf k}$).
However, our present interest is to include this renormalization
effect for an elliptical Fermi surface system and gap
with a possible $s$-component appropriate to an orthorhombic system.
The Green's function (\ref{eq:G}) and above self-energies are solved 
self-consistently along with the gap equation

\begin{eqnarray}
\Delta_{\bf k}={T\over 2}\sum_{{\bf k^\prime},\omega_n}
g({\bf k,k^\prime}){\rm Tr}
[\hat{\tau}_1 \hat{G}({\bf k^\prime},i\omega_n)],
\label{eq:gap.eq}
\end{eqnarray}
where $g({\bf k,k^\prime})$ is the pairing interaction.
 
\section{Anisotropic Model}
\label{sec:model}

\begin{figure}[h]
\vspace{-0.5cm}
\postscript{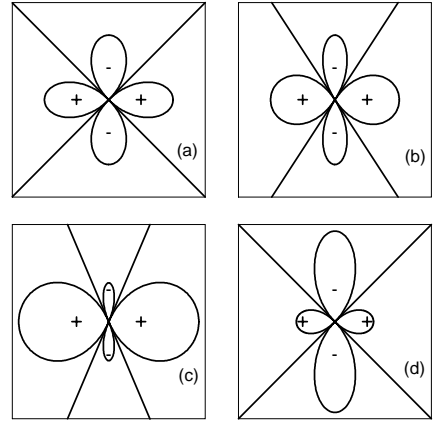}
\vspace{-0.5cm}
\caption{Order parameters on the Fermi surface
given by Eq.~(\protect\ref{eq:gap.p}).
(a): $\alpha=s=0$ (pure $d$-wave gap);
(b): $\alpha=0.4$, $s=0$;
(c): $\alpha=0.4$, $s=0.4$;
(d): $\alpha=0.4$, $s=-0.4$. The straight lines denote the nodal angles.}
\label{fig1}
\end{figure}

We consider a layered superconductor of a $d_{x^2-y^2}$+$s$-wave 
order parameter
 
\begin{eqnarray}
\Delta_{\bf k}= \Delta_0(\hat{k}_x^2-\hat{k}_y^2+s),
\label{eq:gap}
\end{eqnarray}
where the constant $s$ denotes the $s$-wave component
which is small but can be {\em positive} or {\em negative} in general.
For YBCO which is orthorhombic, 
the band structure is chosen to be elliptical 
 
\begin{eqnarray}
\xi(k_x,k_y)={\hbar^2 k_x^2\over 2m_x}+
{\hbar^2 k_y^2\over 2m_y}-\epsilon_{F},
\label{eq:ellipse.band}
\end{eqnarray}
where $\epsilon_{F}$ is the Fermi energy and
taking into account the CuO chain along the
$k_y$ axis, the effective mass $m_x>m_y$.
Our approach follows Kim and Nicol \cite{KN95} closely.
It is convenient to use a transformation
 
\begin{eqnarray}
k_\mu\equiv p_\mu \sqrt{2 m_\mu\over m_x +m_y}~;~~~~~~~~~~~~(\mu=x,y)~
\label{eq:transform}
\end{eqnarray}
and consequently Eqs.~(\ref{eq:gap}) and (\ref{eq:ellipse.band}) are 
transformed from the ${\bf k}$ frame to ${\bf p}$ frame according to
 
\begin{eqnarray}
\Delta_{\bf k}\rightarrow \Delta_{\bf p}= \Delta_0 f(\phi)
\label{eq:gap.p}
\end{eqnarray}
with

\begin{eqnarray}
f(\phi)={\cos(2\phi)+\alpha
\over 1+\alpha\cos(2\phi)}+s~~;~~~~~
\alpha\equiv {m_x - m_y\over m_x + m_y}>0
\label{eq:f}
\end{eqnarray}
and
 
\begin{eqnarray}
\xi(k_x,k_y)\rightarrow
\xi(p_x,p_y)={\hbar^2 \over m_x+m_y}(p_x^2+p_y^2)-\epsilon_{F}.
\label{eq:ellipse.band.p}
\end{eqnarray}
The denominator in (\ref{eq:f}) was ignored in the work of Kim and
Nicol \cite{KN95}. In (\ref{eq:gap.p}), $\phi$
denotes the azimuthal angle in the ${\bf p}$ frame. 
The cylindrical Fermi surface case corresponds to $m_x=m_y$ or $\alpha=0$. 
It is worth noting that the gap given in (\ref{eq:gap.p}) viewed from
the ${\bf p}$ frame is not a simple $d$+$s$-wave gap, $\Delta_{\bf p}= 
\Delta_0 (\cos(2\phi)+s)$, unless $\alpha=0$.
Fig.~\ref{fig1} displays the gaps given by (\ref{eq:gap.p}) and (\ref{eq:f})
with various choice of $\alpha$ and $s$.
The general feature of these pictures is that
the gap nodes are shifted off the diagonals
and for the case $\alpha=-s$, the nodes are push back to the diagonal.
These are to be compared by the pure $d$-wave case ($\alpha=s=0$) case.

The momentum ${\bf k}$ sum can be transformed to ${\bf p}$ sum
which in turn can be replaced by an integration
 
\begin{eqnarray}
\sum_{\bf k}\rightarrow \sum_{\bf p}=
2N(0)\int_{-\infty}^\infty d\xi_{\bf p}\int_0^{2\pi}
{d\phi\over 2\pi},
\label{eq:mom.sum}
\end{eqnarray}
with $N(0)=(m_x +m_y)/4\pi\hbar^2$ defined as the
density of states per spin on the  Fermi surface in ${\bf p}$ frame.
In the normal state in which ${\Delta}_{\bf k}=0$ and hence $G_1=\Sigma_1=0$,
the only nontrivial self-energy in (\ref{eq:Sigmas.1})
is $\Sigma_0=n_i G_0/(c^2- G_0^2)$.
Using (\ref{eq:mom.sum}), one can easily work out 
the (isotropic) scattering rate
$i\Sigma_0\equiv (1/2\tau)$, where
in the Born scattering ($c\gg 1$) limit,
$1/2\tau=2\pi N(0)n_i v_i^2$, while in the
resonant scattering ($c\ll 1$) limit,  $1/2\tau=n_i/2\pi N(0)$.

Using the spectral representation for the imaginary frequency
Green's function, analytically continuing to real frequency
from imaginary frequency ($i\omega_n\rightarrow \omega+i\delta$),
and then performing the frequency sum and
momentum sum gives the real part of the Kubo function

\begin{eqnarray}
K^\prime_{\mu \nu}(\Omega)&=&
{N(0)e^2\over 2}\int_0^{2\pi} {d\phi\over 2\pi}\gamma_\mu(\phi)
\gamma_\nu(\phi) \int_{-\infty}^\infty d\omega
\nonumber\\
&\times& \left\{\left[f(\omega)-f(\omega-\Omega)\right]
{\rm Re} I_{+-}(\Omega,\omega)\right.\nonumber\\
&-& \left.\left[f(\omega)+f(\omega-\Omega)\right]
{\rm Re} I_{++}(\Omega,\omega)\right\}
\label{eq:K1}
\end{eqnarray}
and the imaginary part of the Kubo function

\begin{eqnarray}
&&K^{\prime\prime}_{\mu \nu}(\Omega)=
-{N(0)e^2\over 2}\int_0^{2\pi} {d\phi\over 2\pi}\gamma_\mu(\phi)
\gamma_\nu(\phi)
\int_{-\infty}^\infty d\omega\nonumber\\
&&\times[f(\omega)-f(\omega-\Omega)]
\left[
{\rm Im} I_{+-}(\Omega,\omega)+{\rm Im} I_{++}(\Omega,\omega) \right].
\label{eq:K2}
\end{eqnarray}
In Eqs.~(\ref{eq:K1}) and (\ref{eq:K2}),
the vertices calculated on the Fermi surface 
are $\gamma_x(\phi)=\sqrt{2\epsilon_F/m_x}\cos\phi$ and
$\gamma_y(\phi)=\sqrt{2\epsilon_F/m_y}\sin\phi$ and we have defined

\begin{eqnarray}
I_{++}(\Omega,\omega)&=&
-{1\over \xi_+}+{\tilde{\omega}_+^\prime(\tilde{\omega}_+
+\tilde{\omega}_+^\prime)+
\tilde{\Delta}_+^\prime(\tilde{\Delta}_+
-\tilde{\Delta}_+^\prime)\over
({\xi}_+ +{\xi}_+^\prime){\xi}_+
{\xi}_+^\prime},
\nonumber\\
I_{+-}(\Omega,\omega)&=&
{1\over \xi_+}+{\tilde{\omega}_-^\prime(\tilde{\omega}_+
+\tilde{\omega}_-^\prime)+
\tilde{\Delta}_-^\prime(\tilde{\Delta}_+
-\tilde{\Delta}_-^\prime)\over
({\xi}_+ -{\xi}_-^\prime){\xi}_+
{\xi}_-^\prime},
\label{eq:I.def}
\end{eqnarray}
with $\tilde{\omega}_\pm\equiv i\tilde{\omega}_n(\omega\pm i\delta)$;
$\tilde{\omega}_\pm^\prime\equiv i\tilde{\omega}_n(\omega-\Omega\pm i\delta)$,
$\tilde{\Delta}_\pm\equiv \tilde{\Delta}(\omega\pm i\delta)$;
$\tilde{\Delta}_\pm^\prime\equiv \tilde{\Delta}(\omega-\Omega\pm i\delta)$,
and ${\xi}_\pm\equiv {\rm sgn}(\omega)\sqrt{\tilde{\omega}_\pm^2
-\tilde{\Delta}_{\pm}^2}$; ${\xi}_\pm^\prime\equiv {\rm sgn}(\omega-\Omega)
\sqrt{(\tilde{\omega}_\pm^\prime)^2-(\tilde{\Delta}_{\pm}^\prime)^2}$ 
which are chosen to have branch cuts such that
${\rm Im}\xi_+, {\rm Im}\xi_+^\prime>0$ and
${\rm Im}\xi_-, {\rm Im}\xi_-^\prime<0$.
The results of Eq.~(\ref{eq:K1})-(\ref{eq:I.def})
were obtained before by Hirschfeld {\em et al.} \cite{HPS} and many others.
The microwave conductivity is given directly by $K_{\mu\nu}^{\prime\prime}$ 
as

\begin{eqnarray}
\sigma_{\mu \nu}(T)= \lim_{\Omega\rightarrow 0}
-{K^{\prime\prime}_{\mu \nu}(\Omega,T)\over \Omega},
\label{eq:microwave}
\end{eqnarray}
while the London penetration depth is given via $K^{\prime}_{\mu\nu}$ as

\begin{eqnarray}
{1\over \lambda_{\mu \nu}^{2}(T)}={4\pi \over c^2}\left[
K^\prime_{\mu \nu}(\Omega=0, T)-K^{\prime(n)}_{\mu \nu}(\Omega=0, T)\right],
\label{eq:lambda}
\end{eqnarray}
with $c$ the speed of light.

\section{Anisotropic Universal Transports}
\label{sec:transport}

\subsection{Microwave Conductivity}
\label{sec:sigma}
 
We consider the zero-temperature limit of the
microwave conductivity. When $T\rightarrow 0$,
$[f(\omega)-f(\omega-\Omega)]/\Omega\approx \partial f(\omega)/
\partial \omega\approx -\delta(\omega)$ as $\Omega\rightarrow 0$.
Consequently from (\ref{eq:K2}) and (\ref{eq:microwave}), we have
 
\begin{eqnarray}
\sigma_{\mu\nu}(0)=N(0)e^2\left\langle
{\gamma_\mu(\phi)\gamma_\nu(\phi)
\gamma^2\over [\gamma^2+\tilde{\Delta}^2(\phi)]^{3\over 2}}\right\rangle,
\label{eq:sigma0}
\end{eqnarray}
where $\langle\cdot\cdot\cdot\rangle$ denotes an
average over the Fermi surface. 
In (\ref{eq:sigma0}), $\gamma= -i\tilde{\omega}(\omega=0)=
i\Sigma_0(\omega=0)$ and $\tilde{\Delta}(\phi)\equiv
\tilde{\Delta}(\phi,\omega=0)=\Delta_0[f(\phi)+\delta]$ 
with $\delta=\Sigma_1(\omega=0)/\Delta_0$, respectively
are the impurity-renormalized 
effective scattering rate and gap at zero frequency.  
For $\delta\ll 1$ and $\alpha+s < 1$ of interest
[see Eq.~(\ref{eq:f})],
it is guaranteed that there exists nodes for the renormalized gap.
The appearance of universal transport in unconventional superconductors
is intimately connected to the presence of nodes in the renormalized gap.
When the anisotropies vanish ($\alpha=s=0$), 
$f(\phi)=\cos(2\phi)$ and $\delta=0$ which
are correct whatever concentration of impurity and
correspond to the unrenormalized pure $d$-wave
gap in a cylindrical Fermi surface case.
We will discuss in the next section, the self-consistent results for 
$\gamma$ and $\delta$ in the two different scattering limits,
Born and resonant.

In the case $\bar{\gamma}\equiv\gamma/\Delta_0 \ll 1$, 
the major contribution to the average in Eq.~(\ref{eq:sigma0})
comes from the small angular area around the nodes of the renormalized gap.
One can then carry out the integration in (\ref{eq:sigma0}) to get
 
\begin{eqnarray}
\sigma_{\nu\nu}(0)\simeq {4N(0)e^2\over \pi \Delta_0}
\left|{\partial f(\phi)\over \partial \phi}\right|^{-1}_{\phi=\phi_0}
\gamma_\nu(\phi_0)\gamma_\nu(\phi_0),
\label{eq:sigma0.ellipse2}
\end{eqnarray}
where $\phi_0$ corresponds to the angle of nodes, {\em i.e.},
$f(\phi_0)=-\delta$.
The cross term $\sigma_{xy}(0)=0$ because of the Fermi-surface average.
More explicitly, Eq.~(\ref{eq:sigma0.ellipse2}) gives 

\begin{eqnarray}
\sigma_{xx}(0)&=&{2e^2\epsilon_{F}N(0)\over \pi \Delta_0 m_x}
{(1+\alpha\eta)^2(1+ \eta)\over \sqrt{1-\eta^2}(1-\alpha^2)},
\nonumber\\
\sigma_{yy}(0)&=&{2e^2\epsilon_{F}N(0)\over \pi \Delta_0 m_y}
{(1+\alpha\eta)^2(1- \eta)\over \sqrt{1-\eta^2}(1-\alpha^2)},
\label{eq:sigma0.ellipse3}
\end{eqnarray}
where we have defined $\eta\equiv\cos(2\phi_0)=
-(\alpha+s+\delta)/[1+\alpha(s+\delta)]$. 
One can easily verify that when $\alpha=s=0$ and hence $\delta=0$ 
corresponding to a $d_{x^2-y^2}$-wave gap in the cylindrical
Fermi surface case ($m_x=m_y$), 

\begin{eqnarray}
\sigma_{xx}(0)=\sigma_{yy}(0)={ne^2\over \pi\Delta_0 m},
\label{eq:sigma.isotropic}
\end{eqnarray}
with $n=k_F^2/2\pi$ for the electronic density.
The universal feature of (\ref{eq:sigma.isotropic}) was first predicted
by Lee \cite{Lee93}.

While the result in Eq.~(\ref{eq:sigma0.ellipse3}) 
does not depend on $\gamma$, it is explicitly dependent of
the gap shift $\delta$ which in turn, is associated with
the impurity concentration. 
In the large resulting anisotropy case 
$[(\alpha+s)^2\gg 1/(\tau\Delta_0)]$ of primary interest (provided the
impurity concentration is low), one can ignore $\delta$ (compared to
$\alpha$ and $s$) in Eq.~(\ref{eq:sigma0.ellipse3}), 
valid for both Born and unitary limits (see Sec.~\ref{sec:self-energy}).  
As a consequence,  
 
\begin{eqnarray}
\sigma_{xx}(0)&=&{2e^2\epsilon_{F}N(0)\over \pi \Delta_0}{A_{xx}\over m_x}
\equiv{\Omega_{{\rm p},xx}^2\over 4\pi}{A_{xx}\over \pi\Delta_0}
\nonumber\\
\sigma_{yy}(0)&=&{2e^2\epsilon_{F}N(0)\over \pi \Delta_0}{A_{yy}\over m_y}
\equiv{\Omega_{{\rm p},yy}^2\over 4\pi}{A_{yy}\over \pi\Delta_0}~,
\label{eq:sigma0.ellipse4}
\end{eqnarray}
where we have defined

\begin{eqnarray}
A_{xx}&=&{1-\alpha^2 \over (1+\alpha s)^2}{(1-\alpha-s+\alpha s)\over
\sqrt{1-\alpha^2-s^2+\alpha^2 s^2}},\nonumber\\
A_{yy}&=&
{1-\alpha^2 \over (1+\alpha s)^2}{(1+\alpha+s+\alpha s)\over
\sqrt{1-\alpha^2-s^2+\alpha^2 s^2}}~~.
\label{eq:A.ab}
\end{eqnarray}
In (\ref{eq:sigma0.ellipse4}), $\Omega_{{\rm p},xx}(\Omega_{{\rm p},yy})$
is the plasma frequency in the $x(y)$ direction \cite{Tanner,TT92}.
In comparison with experiment, the plasma frequency is a measured
quantity and so $\sigma_{ii}(0)$ is a measure of $A_{ii}/\Delta_0$.
Note the extra factor of $A_{ii}$ in (\ref{eq:sigma0.ellipse4})
as compared with the well-known result (\ref{eq:sigma.isotropic})
which is $\sigma(0)=(\Omega_{\rm p}^2/4\pi)(1/\pi\Delta_0)$.

Eq.~(\ref{eq:sigma0.ellipse4})
represents the anisotropic universal feature in
the microwave conductivity for an $d_{x^2-y^2}$+$s$-wave superconductor
with an orthorhombic band structure.
The ratio

\begin{eqnarray}
{\sigma_{xx}(0) \over \sigma_{yy}(0)}={\Omega^2_{{\rm p},xx}\over 
\Omega^2_{{\rm p},yy}}{A_{xx}\over A_{yy}}
\label{eq:ratio}
\end{eqnarray}
should have some experimental consequence.
We note that the low-temperature
correction terms to $\sigma_{ii}(0)$ will be proportional to
$T^2/\gamma^2$ with $\gamma$ the effective impurity scattering rate. 
Thus when $T\ll \gamma$, the anisotropic universal values can be attained.
 
In the small anisotropy case $[(\alpha+s)^2\ll 1/(\tau\Delta_0)]$, 
while $\alpha$ and $|s|$ may be
large individually, it can be shown that
$\delta\sim (\alpha+s)\sim\eta$ are all small 
(see Sec.~\ref{sec:self-energy}).  Consequently, 
one ignores $\eta$ in (\ref{eq:sigma0.ellipse3}) to obtain

\begin{eqnarray}
\sigma_{xx}(0)&=&{\Omega_{{\rm p},xx}^2\over 4\pi^2 \Delta_0 (1-\alpha^2)},
\nonumber\\
\sigma_{yy}(0)&=&{\Omega_{{\rm p},yy}^2\over 4\pi^2 \Delta_0 (1-\alpha^2)}.
\label{eq:sigma0.ellipse5}
\end{eqnarray}
A ratio $\sigma_{xx}(0)/\sigma_{yy}(0)=\Omega_{{\rm p},xx}^2/
\Omega_{{\rm p},yy}^2$ is given in this limit.
In the case of $\alpha\rightarrow 0$, $\sigma_{xx}(0)=\sigma_{yy}(0)$ in
(\ref{eq:sigma0.ellipse5}) and
they reduce to Eq.~(\ref{eq:sigma.isotropic}).
 
\subsection{London Penetration Depth}
\label{sec:lambda}
 
We consider here, the {\em anisotropic} London penetration depths within
our anisotropic Fermi surface model together with
a $d_{x^2-y^2}$+$s$-wave order parameter.
The results are to be compared to the recent work of Kim and Nicol \cite{KN95}
who effectively have considered the
anisotropic penetration depth with $\alpha=0$
because they ignore the denominator in (\ref{eq:f}).
In the pure limit, using Eqs.~(\ref{eq:K.def}) and (\ref{eq:lambda}) 
it can be shown that 
the square of the inverse penetration depth is given by
 
\begin{eqnarray}
{1\over \lambda_{\nu\nu}^{2}}(T)={8\pi e^2\over c^2}{1\over \Omega}
\sum_{\bf k} v_{{\bf k},\nu}^2\left[
{\partial f(E_{\bf k})\over\partial E_{\bf k}}-
{\partial f(\epsilon_{\bf k})\over\partial \epsilon_{\bf k}} \right]
\label{eq:pene.depth}
\end{eqnarray}
and is associated with the superfluid density.
In the low-temperature limit, one obtains relatively simple results
 
\begin{eqnarray}
{1\over \lambda_{xx}^{2}(T)}&=&{\Omega_{{\rm p},xx}^2\over c^2}
\left[1-2\ln 2 \left({k_B T\over \Delta_0}\right)
A_{xx}\right],\nonumber\\
{1\over \lambda_{yy}^{2}(T)}&=&{\Omega_{{\rm p},yy}^2\over c^2}
\left[1-2\ln 2 \left({k_B T\over \Delta_0}\right)
A_{yy}\right],
\label{eq:pene.depth.xy.ab}
\end{eqnarray}
where $A_{xx}$ and $A_{yy}$ are given in Eq.~(\ref{eq:A.ab}).

Our result confirms Kim and Nicol's result when we set $\alpha=0$.
The linear in temperature behavior in Eq.~(\ref{eq:pene.depth.xy.ab})
is expected for an unconventional
superconductor with gap nodes on Fermi surface in the clean limit.
When impurities are present,
one instead expects a $T^2$ law for the low-temperature
penetration depth (see Ref.~\cite{KN95}).
In view of (\ref{eq:pene.depth.xy.ab}), the discrepancy exhibited in
$a$ and $b$-axis penetration depths can be studied in two ways.
First, the zero-temperature anisotropic penetration depth ratio
 
\begin{equation}
{\lambda_{yy}^{2}(0)\over \lambda_{xx}^{2}(0)}={\Omega_{{\rm p},xx}^2
\over \Omega_{{\rm p},yy}^2}={m_y\over m_x}=
{1-\alpha\over 1+\alpha}.
\label{eq:lambda0.ratio}
\end{equation}
Secondly, the low-temperature slope ratio
 
\begin{equation}
{\left.d[\lambda_{xx}^{2}(0)/\lambda_{xx}^{2}(T)]/dT\right|_{T\rightarrow 0}
\over 
\left.d[\lambda_{yy}^{2}(0)/\lambda_{yy}^{2}(T)]/dT\right|_{T\rightarrow 0}}
={A_{xx}\over A_{yy}}.
\label{eq:lambda_slope.ratio}
\end{equation}
To fit the experimental data of low-temperature penetration depth
by Bonn {\em et al.} \cite{Bonn} and the normal-state resistivity 
by Zhang {\em et al.} \cite{Zhang}, 
it requires $\alpha=0.4$ in (\ref{eq:lambda0.ratio}) and
then a value of $s=-0.25$ is obtained using
(\ref{eq:lambda_slope.ratio}) \cite{Ewald}.

\begin{figure}[h]
\vspace{-0.5cm}
\postscript{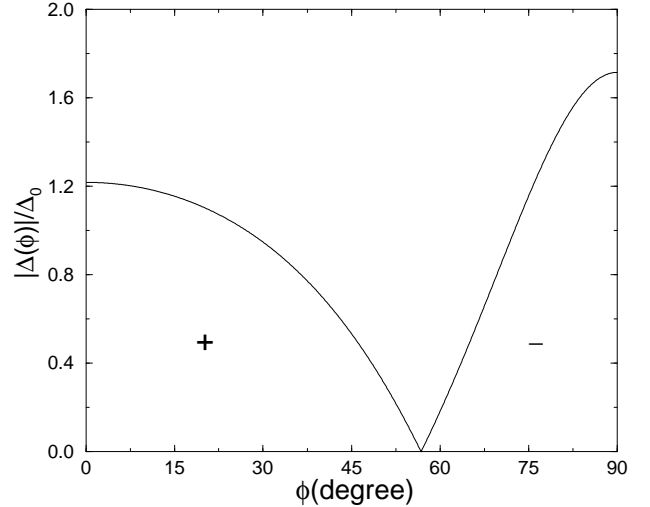}
\vspace{-0.5cm}
\caption{Relative magnitude of order
parameter [given by Eq.~(\protect\ref{eq:gap.p})]
vs. angle using parameters fit to penetration depth low-temperature slopes
($\alpha=0.4$ and $s=-0.25$).}
\label{fig2}
\end{figure}

In Fig.~\ref{fig2}, we plot the relative size of
gap $\Delta(\phi)$ as a function of angle using parameters
fit to low-temperature
penetration depth slopes (alpha = 0.4, s = -0.25). As shown in
Fig.~\ref{fig2}, the gap node is shifted from $45^{\rm o}$ and the overall
gap magnitude is highly anisotropic. These may have observable consequence
in future ARPES experiments.

\subsection{Thermal Conductivity}
\label{sec:kappa}

In analogy to Eq.~(\ref{eq:K2}) which gives the
microwave conductivity, the thermal conductivity is given by

\begin{eqnarray}
\kappa_{\mu \nu}(T)&=&-
{N(0)\over 2T}\int_0^{2\pi} {d\phi\over 2\pi}\gamma_\mu(\phi)
\gamma_\nu(\phi)
\int_{-\infty}^\infty d\omega~ \omega^2
{\partial f(\omega)\over \partial\omega }
\nonumber\\
&\times& \left[
{\rm Im} I_{+-}(\Omega=0,\omega)+{\rm Im} I_{++}(\Omega=0,\omega) \right].
\label{eq:thermoconductivity}
\end{eqnarray}
As we seek the anisotropic universality in thermal conductivity,
we are specifically interested in the $T\rightarrow 0$ regime
in which the inelastic scattering is unimportant.
When $T\rightarrow 0$, the $\omega$ integration in 
Eq.~(\ref{eq:thermoconductivity})
is limited to $\omega\approx 0$ and consequently

\begin{eqnarray}
{\kappa_{\mu \nu}(T\rightarrow 0)\over \sigma_{\mu \nu}(T\rightarrow 0)}
&=&{1\over T}\int_{-\infty}^{\infty}d\omega~ \omega^2
{\partial f(\omega)\over \partial \omega}\left/
e^2\int_{-\infty}^{\infty}dx~
{\partial f(\omega)\over \partial \omega}\right.\nonumber\\
&=& {\pi^2\over 3}\left({k_B\over e}\right)^2 T.
\label{eq:WF.law}
\end{eqnarray}
Therefore the Wiedemann-Franz law (or Sommerfeld law)
is satisfied. The above result is
valid for both $a$ and $b$ directions.
Previously Graf {\em et al.} \cite{GYSR96} have shown
that for an isotropic Fermi surface band structure,
the thermal conductivity saturates and
obeys the Wiedemann-Franz law at low temperatures ($T\ll \gamma$) for
disordered gapless superconductors.  
As emphasized by Graf {\em et al.} \cite{GYSR96},
similar to the condition of universal microwave conductivity,
the validity of the Wiedemann-Franz law at $T\rightarrow 0$ is
a unique feature of gapless superconductors in which the impurity
induces a finite density of states at the zero energy.
Our predicted
{\em anisotropic} universal thermal conductivity can be written explicitly
as [when $(\alpha+s)^2\agt 1/(\tau\Delta_0)$]

\begin{eqnarray}
\left.{\kappa_{xx}(T)\over T}\right|_{T\rightarrow 0}
&=&{1\over 12}\left({k_B\over e}\right)^2 \Omega_{{\rm p},xx}^2
{A_{xx}\over \Delta_0}
\nonumber\\
\left.{\kappa_{yy}(T)\over T}\right|_{T\rightarrow 0}
&=&{1\over 12}\left({k_B\over e}\right)^2 \Omega_{{\rm p},yy}^2
{A_{yy}\over \Delta_0} ~.
\label{eq:K.ellipse4}
\end{eqnarray}
Thus, similar to Eq.~(\ref{eq:ratio}), 

\begin{eqnarray}
{\kappa_{xx}(0) \over \kappa_{yy}(0)}={\Omega_{{\rm p},xx}^2
\over \Omega_{{\rm p},yy}^2}{A_{xx}\over A_{yy}}.
\label{eq:kappa.ratio}
\end{eqnarray}

Recently, an experiment done by Taillefer {\em et al.}
\cite{TLGBA97} which measures the in-plane low-temperature thermal
conductivity of YBa$_2$Cu$_3$O$_{6.9}$ at different Zn substitutions for Cu
has confirmed the universal feature in thermal conductivity.
More recently, Chiao {\em et al.} \cite{Taillefer}  
have observed the anisotropic universal 
thermal conductivity in YBCO.
They found that the thermal conductivity 
saturates at low temperatures both along $a$ and $b$ directions from which
an anisotropy $\kappa_b(T\rightarrow 0)/\kappa_a(T\rightarrow 0)=1.3$ 
is obtained \cite{Taillefer}.
If we take $\alpha=0.4$ as predicted by anisotropic in-plane
penetration depth
\cite{Bonn} and normal-state resistivity \cite{Zhang} data, it would imply
a negative but large $s=-0.60$ in use of Eq.~(\ref{eq:kappa.ratio}). 
Unfortunately such a large value of $s$ reverses the order of the slopes
in the penetration depth making the $x$ direction steeper than the $y$
direction in contradiction to experiment.
For very small $1/\tau$, it can also imply a very small value of $\gamma$ 
so that saturation effects will be seen only at very low temperature again 
in contradiction to experiment.
On the other hand, with $\alpha=0.4$ and $s=-0.60$, the value of 
$A_{xx}\simeq 1.9$ makes it unnecessary to invoke a factor of 2 in the
slope of the gap at the nodes needed in the
analysis of Taillefer {\em et al.} \cite{TLGBA97} who
concluded that the actual gap grows out of zero a factor of 2 more
slowly than in a pure $d$-wave case.

The large negative value of $s$ needed to explain the observed
small anisotropy between $a$ and $b$ direction in the
thermal conductivity is not believed to be           
physical. A more likely explanation may be in the
observation that even for pure crystals of optimally doped
YBCO, there is a large residual resistivity on the
chains and none on the planes. This highly anisotropic residual resistivity
has not been accounted for in our work.

\section{Self-energies and density of states}
\label{sec:self-energy}

We discuss here the self-consistent results for 
the dimensionless self-energies $\bar{\gamma}$ and $\delta$.
One recalls that $\bar{\gamma}=i\Sigma_0(\omega=0)/\Delta_0$
and $\delta=\Sigma_1(\omega=0)/\Delta_0$.
The self-energies $\Sigma_i$ in (\ref{eq:Sigmas.1}) are comprised of
the integrated Green's functions $G_i$ which are given by

\begin{eqnarray}
G_0(\omega)=-2i\pi N(0)\left\langle
{\tilde{\omega} \over \sqrt{\tilde{\omega}^2-\tilde{\Delta}^2(\phi,\omega)}}
\right\rangle
\label{eq:G0}
\end{eqnarray}
and

\begin{eqnarray}
G_1(\omega)=-2i\pi N(0)\left\langle
{\tilde{\Delta}(\phi,\omega) \over 
\sqrt{\tilde{\omega}^2-\tilde{\Delta}^2(\phi,\omega)}}
\right\rangle,
\label{eq:G1}
\end{eqnarray}
after the energy integration is done.
For the case $\delta,\bar{\gamma}\ll 1$ and $|\alpha+s| <1$
of interest (in which the gap is guaranteed to have nodes), 
we have found to leading order,

\begin{eqnarray}
&&G_0(\omega=0)\simeq 4iN(0)\bar{\gamma}\ln \bar{\gamma}\nonumber\\
&&G_1(\omega=0)\simeq -4N(0)(\alpha+s+\delta).
\label{eq:G01,w=0}
\end{eqnarray}

\subsection{Born Limit}
\label{sec:born}

Using Eq.~(\ref{eq:G01,w=0}) in (\ref{eq:Sigmas.1}),
in the Born limit ($c\gg 1$), we obtain

\begin{eqnarray}
\bar{\gamma}\simeq e^{-\pi\tau\Delta_0}~~;~~~~~~
\delta\simeq {\alpha+s\over \pi\tau\Delta_0}.
\label{eq:Born}
\end{eqnarray}
Typically $\tau\Delta_0\sim 100$ (providing the impurity
concentration is low), one thus retains $\bar{\gamma}\ll 1$ and 
$\delta\ll |\alpha+s| <1$. The latter validates Eqs.~(\ref{eq:sigma0.ellipse4})
and (\ref{eq:ratio}). 

\subsection{Unitary Limit}
\label{sec:unitary}

In the unitary limit ($c\ll 1$) of primary interest here, 
and when the resulting anisotropy is large such that
$(\alpha+s)^2\gg 1/(\tau\Delta_0)$, we found

\begin{eqnarray}
\bar{\gamma}\simeq e^{-4\tau\Delta_0(\alpha+s)^2/\pi}~~;~~~~~~
\delta\simeq {\pi\over 4\tau\Delta_0(\alpha+s)}.
\label{eq:Unitary.b.alpha}
\end{eqnarray}
Clearly Eq.~(\ref{eq:Unitary.b.alpha}) gives
$\bar{\gamma},\delta\ll 1$.
Thus one can ignore $\delta$ in (\ref{eq:sigma0.ellipse3})
to get Eqs.~(\ref{eq:sigma0.ellipse4}) and (\ref{eq:ratio}).
Eq.~(\ref{eq:Unitary.b.alpha}) indicates that
the effective scattering rate is strongly
suppressed compared to isotropic case.
When the resulting anisotropy is small [$(\alpha+s)^2\alt 1/(\tau\Delta_0)$],
though $\alpha$ and $|s|$ may be large individually,
the system undergoes a crossover from an anisotropic regime to a 
``quasi-isotropic'' regime in which the 
gap shift $\delta$ is small and is linear in
and about the size of $(\alpha+s)$.
When $\alpha=-s$, $\delta$ vanishes as for a pure $d$-wave
superconductor in an isotropic Fermi surface.
More explicitly, to second order in $(\alpha+s)$, we find

\begin{eqnarray}
&&\bar{\gamma}\simeq \left({-\pi\over 4\tau\Delta_0 \ln\bar{\gamma}_0}
\right)^{1\over 2}
\left[1+{2\tau\Delta_0\ln\bar{\gamma}_0(\alpha+s)^2\over
\pi(\ln\bar{\gamma}_0+1)^2}\right]\nonumber\\
&& \delta\simeq -{\alpha+s\over 1+\ln\bar{\gamma}_0},
\label{eq:Unitary.s.alpha}
\end{eqnarray}
where $\bar{\gamma}_0$ satisfies $\bar{\gamma}_0^2\ln\bar{\gamma}_0=-
\pi/(4\tau\Delta_0)$. Typically $\ln\bar{\gamma}_0<-2$ and
with $(\alpha+s)^2\ll 1/(\tau\Delta_0)$, 
the second term in the 
expression $\bar{\gamma}$ in Eq.~(\ref{eq:Unitary.s.alpha}) is small.
One thus obtains a $\gamma$ value similar to that
for the case of a pure $d$-wave
superconductor with an isotropic Fermi surface.

One should keep in mind that while when $(\alpha+s)^2\ll 1/(\tau\Delta_0)$,
the results for
$\gamma$ and $\delta$ are similar to those of an isotropic system,
the universal microwave conductivities are still
{\em anisotropic}, as given by Eq.~(\ref{eq:sigma0.ellipse5}).
They reduce to the universal value of Eq.~(\ref{eq:sigma.isotropic})
only when $\alpha=0$.

\begin{figure}[h]
\vspace{-0.5cm}
\postscript{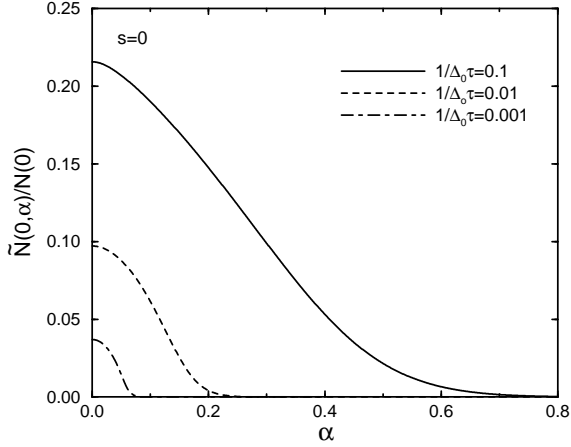}
\vspace{-0.5cm}
\caption{Plot of zero-frequency density of states
$\tilde{N}(0,\alpha)$ scaled to $N(0)$, in the unitary limit,
as a function of $\alpha$ for different
choices of scattering rate $1/\Delta_0 \tau$. Here we choose $s=0$.}
\label{fig3}
\end{figure}

\subsection{Density of States}
\label{sec:dos}

\begin{figure}[h]
\vspace{-0.5cm}
\postscript{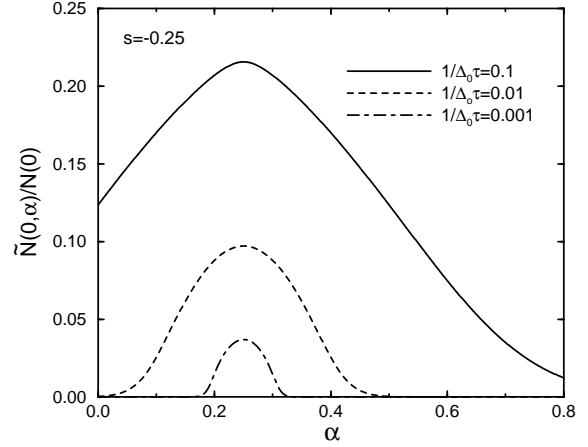}
\vspace{-0.5cm}
\caption{Similar plot as Fig.~\protect\ref{fig3} with $s=-0.25$.}
\label{fig4}
\end{figure}

Using the result of (\ref{eq:G01,w=0}), the
impurity renormalized density of states at zero frequency is given by
 
\begin{eqnarray}
\tilde{N}(0)=-{i\over 2\pi}G_0(\omega=0)=
-{2N(0)\over \pi}\bar{\gamma}\ln\bar{\gamma},
\label{eq:DOS}
\end{eqnarray}
which is dependent on the impurity concentration (or $1/\tau$)
and the magnitudes of $\alpha$ and $s$.
In Fig.~\ref{fig3}, we compute and plot the 
$\alpha$ dependence of $\tilde{N}(0)$, in the unitary scattering limit,
for three different values of $1/(\tau\Delta_0)$,
taking $s=0$. Similar plot is made in Fig.~\ref{fig4} using $s=-0.25$.
As already predicted for the effective 
scattering rate $\gamma$, the density of states
is strongly suppressed when $(\alpha+s)^2\agt 1/(\tau\Delta_0)$.
For example, in case of $1/(\tau\Delta_0)\sim 0.01$,
$\tilde{N}(0)$ is suppressed by $50\%$ when $(\alpha+s)\agt 0.1$ 
compared to $(\alpha+s)=0$ case (see Figs.~\ref{fig3} and \ref{fig4}).
In comparison with experimental results,
the large discrepancy between $a$ and $b$-axis
penetration depths \cite{Bonn} and normal-state resistivity \cite{Zhang}
suggest that a large $\alpha\sim 0.4$ is required
so that $\tilde{N}(0)$ is strongly suppressed.
On the other hand, the low but finite-temperature
universal observed value of the
thermal conductivity \cite{TLGBA97}
suggests that the effective scattering rate or
density of states at zero frequency is not strongly suppressed.
For one to be consistent with the other, it implies that
a large $\alpha$ is accompanied by a large but negative $s$
which reduces the effect of $\alpha$ and
leads to less suppression of the density of states at zero frequency.

\subsection{$T_c$ Suppression}
\label{sec:tc}

In the standard approach, we choose the pairing interaction
to have a separable form
$g({\bf k,k^\prime})=-gf(\phi)f(\phi^\prime)$ ($g>0$),
where $f(\phi)$ is given in (\ref{eq:f}). This reduces 
the gap equation of Eq.~(\ref{eq:gap.eq}) to the form 

\begin{eqnarray}
{1\over g}=2\pi N(0)T\sum_{\omega_n}^\prime \left\langle
{f(\phi) \tilde{\Delta}(\phi,\omega_n)/\Delta_0\over
\sqrt{\tilde{\omega}_n^2+\tilde{\Delta}^2(\phi,\omega_n)}}\right\rangle,
\label{eq:gap.eq.sim}
\end{eqnarray}
where the prime denotes a cutoff. When $T\rightarrow T_c$, it is
sufficient to use the linearized version of (\ref{eq:gap.eq.sim}),
{\em i.e.}, ignoring the $\tilde{\Delta}$ term in denominator.
In a parallel manner, one can neglect the $G_1^2$ term in the denominators
of the self-energies in (\ref{eq:Sigmas.1}). This enables one to 
obtain, valid for both Born and unitary limits,

\begin{eqnarray}
\tilde{\Delta}(\phi,\omega_n)&=&\Delta_0\left[f(\phi)+{1\over 2\tau}
{\langle f(\phi)\rangle\over |\omega_n|}\right]
\nonumber\\
\tilde{\omega}_n&=&\omega_n+{1\over 2\tau}{\rm sgn}(\omega_n).
\label{eq:se.T=Tc}
\end{eqnarray}
The second line of (\ref{eq:se.T=Tc})
is identical to the solution for the normal state, as expected.
Thus using (\ref{eq:se.T=Tc}), the linearized version of (\ref{eq:gap.eq.sim})
is reduced to

\begin{eqnarray}
{1\over g}=4\pi N(0)T\sum_{n \geq 0}^\prime 
{\displaystyle \langle f^2(\phi)\rangle+ {1\over 2\tau\omega_n}\langle 
f(\phi)\rangle^2 \over \displaystyle \omega_n + {1\over 2\tau}}.
\label{eq:gap.eq.sim.1}
\end{eqnarray}
We have found quite generally

\begin{eqnarray}
\ln\left({T_c\over T_{c0}}\right)&=&
{\langle f^2(\phi)\rangle-\langle f(\phi)\rangle^2 \over
\langle f^2(\phi)\rangle}\nonumber\\
&\times&\left[\psi\left({1\over 2}\right)-
\psi\left({1\over 2}+{1/2\tau\over 2\pi T_c}\right)\right],
\label{eq:Tc.sup}
\end{eqnarray}
where $T_{c0}$ is critical temperature in the pure case.

In our case [see (\ref{eq:f})],

\begin{eqnarray}
\langle f^2(\phi)\rangle&=&{1-\sqrt{1-\alpha^2}\over \alpha^2}(1+2\alpha s)
+s^2\nonumber\\
\langle f(\phi)\rangle&=&{1-\sqrt{1-\alpha^2}\over \alpha}+s
\label{eq:f.avg}
\end{eqnarray}
and to leading order,
$(\langle f^2\rangle-\langle f\rangle^2)/
\langle f^2\rangle=(2-\alpha^2)/[2+4s(\alpha+s)]$.
When $1/2\tau \ll T_{c0}$,
we have obtained using (\ref{eq:Tc.sup})

\begin{eqnarray}
{T_c\over T_{c0}}=1-\left[{2-\alpha^2\over 2+4s(\alpha+s)}\right]
{\pi\over 8\tau T_{c0}},
\label{eq:Tc.sup.1}
\end{eqnarray}
which depends linearly on the impurity concentration.
When $1/2\tau \gg T_{c0}$, we obtain
 
\begin{eqnarray}
{T_c\over T_{c0}}=
\left({1\over 4\pi \tau T_{c0}}
\right)^{-\left[{2-\alpha^2\over 2+4s(\alpha+s)}\right]}.
\label{eq:Tc.sup.2}
\end{eqnarray}
It is shown clearly in Eqs.~(\ref{eq:Tc.sup.1}) and
(\ref{eq:Tc.sup.2}) that,
in contrast to the cylindrical Fermi surface and pure $d$-wave gap case
($\alpha=s=0$) in which $T_c$ is strongly 
suppressed compared to $T_{c0}$, 
the presence of anisotropy (finite $\alpha$ or $s$ case) 
reduces the effect of the suppression of $T_c$.

\section{Conclusions}
\label{sec:conclusions}

In this paper, we have studied the low-temperature 
microwave conductivity and thermal conductivity 
for a $d_{x^2-y^2}$+$s$-wave superconductor with an 
orthorhombic elliptical Fermi surface. Similar 
to the universal behaviors found
in the microwave conductivity \cite{Lee93} and thermal
conductivity \cite{GYSR96} for a $d_{x^2-y^2}$-wave
superconductor with a cylindrical Fermi surface,
{\em anisotropic} universal features are found in the present case.
The effects of Fermi surface orthorhombicity and additional $s$ component 
to the gap on the penetration depth, impurity induced $T_c$ suppression, 
and the zero-frequency density of states are also considered.
It is found that, compared to the cylindrical Fermi surface
and pure $d_{x^2-y^2}$-wave gap case,
a small amount of anisotropy (either band anisotropy or gap
admixture) will lead to a
strong suppression of effective scattering rate 
and thus the density of states at zero frequency.
Nevertheless, experimental data suggests that a large band structure
anisotropy effect is compensated by a large but negative $s$-wave gap
component.
 
\acknowledgments
We thank Ewald Schachinger and
Bo\v{z}idar Mitrovi\'{c} for stimulating discussion
and May Chiao for sending us thermal conductivity results prior to
publication. This work was supported in part by
Natural Sciences and Engineering Research Council (NSERC) of Canada,
Canadian Institute for Advanced Research (CIAR), and
by National Science Council (NSC) of Taiwan under Grant No.
NSC 87-2112-M-003-014.

\end{document}